\documentclass {journal}

\usepackage {a4wide}
\usepackage {amssymb}
\usepackage [usenames] {color}
\usepackage{graphicx}
\usepackage {enumerate}
\usepackage {citesort}

\usepackage [usenames] {color}
\usepackage [left,pagewise] {lineno}

\usepackage{fancyhdr}

\definecolor {infocolor} {rgb} {0.6,0.6,0.6}

\definecolor {sepia} {rgb} {0.75,0.30,0.15}

\everymath{\color{sepia}}

\usepackage {amsmath}
\usepackage {xspace}

\newtheorem {theorem} {Theorem}

\newtheorem {lemma} {Lemma}
\newtheorem {observation} {Observation}
\newtheorem {corollary} {Corollary}

\newtheorem {definition} {Definition}
\newenvironment {proof}{\textbf {Proof:}}{\hfill \ensuremath {\boxtimes}}

\newcommand{\maxno}[1]{\ensuremath{T_{#1}(n)}}
\newcommand{\minno}[1]{\ensuremath{t_{#1}(n)}}

\newcommand{\maxnok}{\maxno{k}}
\newcommand{\minnok}{\minno{k}}

\newcommand{\Pts}{\cal P}

\newcommand {\hodt}{higher order Delaunay triangulation}
\newcommand {\hodts}{higher order Delaunay triangulations}
\newcommand {\Hodts}{Higher order Delaunay triangulations}
\newcommand {\fodt}{first order Delaunay triangulation}
\newcommand {\fodts}{first order Delaunay triangulations}
\newcommand {\Fodts}{First order Delaunay triangulations}

\newcommand {\orderk}{\mbox{order-$k$}}

\newcommand{\bE}{{\mathbb E}}
\newcommand{\bP}{{\mathbb P}}

\graphicspath{{./figures/}}

\begin{document}

\title{On the Number of Higher Order Delaunay Triangulations}

\author{%
  Dieter Mitsche\thanks{Centre de Recerca Matem\`{a}tica, Universitat Aut\`{o}noma de Barcelona,
  \tt{dmitsche@crm.cat}.}
  \and
  Maria Saumell\thanks{Dept. Matem\`{a}tica Aplicada II, Universitat Polit\`{e}cnica de Catalunya,
   \tt{maria.saumell@upc.edu}.}
  \and
  Rodrigo I. Silveira\thanks{Dept. Matem\`{a}tica Aplicada II, Universitat Polit\`{e}cnica de Catalunya,
   \tt{rodrigo.silveira@upc.edu}.}
}

\maketitle

\begin{abstract}
\Hodts\ are a generalization of the Delaunay triangulation which
provides a class of well-shaped triangulations, over which extra
criteria can be optimized. A triangulation is order-$k$ Delaunay
if the circumcircle of each triangle of the triangulation contains
at most $k$ points. In this paper we study lower and upper bounds
on the number of \hodts,  as well as their expected number for
randomly distributed points. We show that arbitrarily large point
sets can have a single \hodt, even for large orders, whereas for
\fodts, the maximum number is $2^{n-3}$. Next we show that
uniformly distributed points have an expected number of at least
$2^{\rho_1 n(1+o(1))}$  \fodts, where $\rho_1$ is an analytically
defined constant ($\rho_1 \approx 0.525785$), and for $k > 1$, the
expected number of order-$k$ Delaunay triangulations (which are
not order-$i$ for any $i < k$) is at least $2^{\rho_k n(1+o(1))}$,
where $\rho_k$ can be calculated numerically.
\end{abstract}

 \section{Introduction}

A triangulation is a decomposition into triangles. In this paper
we are interested in triangulations of point sets in the Euclidean
plane, where the input is a set of points in the plane, denoted
$\Pts$, and a triangulation is defined as a subdivision of the
convex hull of $\Pts$ into triangles whose vertices are the points
in $\Pts$.

It is a well-known fact that $n$ points in the plane can have many different triangulations.
For most application domains, the choice of the triangulation is important, because different triangulations can have different effects.
For example, two important fields in which triangulations are frequently used are finite element methods and terrain modeling.
In the first case, triangulations are used to subdivide a complex domain by creating a mesh of simple elements (triangles), over which a system of differential equations can be solved more easily.
In the second case, the points in $\Pts$ represent points sampled from a terrain (thus each point has also an elevation), and the triangulation provides a bivariate interpolating surface, providing an elevation model of the terrain.
In both cases, the shapes of the triangles can have serious consequences on the result.
For mesh generation for finite element methods, the aspect ratio of the triangles is particularly important, since elements of large aspect ratio can lead to poorly-conditioned systems.
Similarly, long and skinny triangles are generally not appropriate for surface interpolation because they can lead to interpolation from points that are too far apart.

In most applications, the need for \emph{well-shaped}
triangulations is usually addressed by using the \emph{Delaunay}
triangulation. The Delaunay triangulation of a point set $\Pts$ is
defined as a triangulation where the vertices are the points in
$\Pts$ and the circumcircle of each triangle (that is, the circle
defined by the three vertices of each triangle) contains no other
point from $\Pts$. The Delaunay triangulation has many known
properties that make it the most widely-used triangulation. In
particular, there are several efficient and relatively simple
algorithms to compute it, and its triangles are considered
\emph{well-shaped}. This is because it maximizes the minimum angle
among all triangle angles, which implies that its angles are---in
a sense---\emph{as large as possible}. Moreover, when the points
are in general position (that is, when no four points are
cocircular and no three points are collinear) it is uniquely
defined. However, this last property can become an important
limitation if the Delaunay triangulation is suboptimal with
respect to other criteria, independent of the shape of its
triangles, as it is often the case in applications.

\begin{figure}[tb]
\centering
\includegraphics{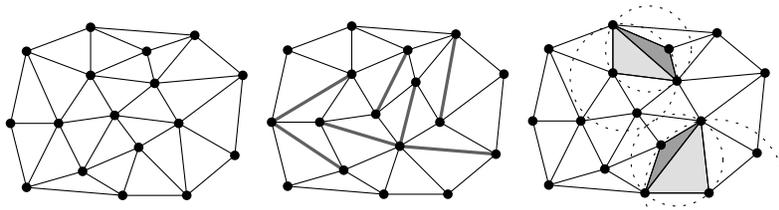}
\caption{Left: A Delaunay triangulation ($k=0$). Center: an
order-$1$ triangulation (with useful-1, non-Delaunay, edges in
gray). Right: an order-$2$ triangulation, with order-1 triangles
in light gray and order-2 triangles in medium gray.}
\label{fig:FigOrders}
\end{figure}

To overcome this limitation, Gudmundsson et al. proposed
\emph{\hodts}~\cite{ghk-hodt-02}. They are a natural
generalization of the Delaunay triangulation that provides
well-shaped triangles, but at the same time,  flexibility to
optimize some extra criterion. They are defined by allowing up to
$k$ points inside the circumcircles of the triangles (see Figure~\ref{fig:FigOrders}). For $k=0$,
each point set in general position has only one higher order
Delaunay triangulation, equal to the Delaunay triangulation. As
the parameter $k$ is increased, more points inside the
circumcircles imply a reduction of the shape quality of the
triangles, but also an increase in the number of triangulations
that are considered. This last aspect makes the optimization of
extra criteria possible, thus providing triangulations that are a
compromise between well-shaped triangles and optimality with
respect to other criteria.

Therefore the importance of \hodts\ lies in multi-criteria
triangulations. Their major contribution is providing a way to
optimize over a---hopefully large---class of well-shaped
triangulations.

A particularly important subclass of \hodts\ are the
\emph{\fodts}, that is, when $k=1$. It has been observed that
already for $k=1$, a point set with $n$ points can have an
exponential number of different triangulations~\cite{s-opt-09}.
This, together with the fact that for $k=1$ the shape of the
triangles is as close as possible to the shape of the Delaunay
triangles (while allowing more than one triangulation to choose
from), make first order Delaunay triangulations especially
interesting. In fact, \fodts\ have been shown to have a special
structure that facilitates the optimization of many
criteria~\cite{ghk-hodt-02}. For example, it has been shown that
many criteria related to measures of single triangles, as well as
some other relevant parameters like the number of local minima,
can be optimized in $O(n \log n)$ time for $k=1$. In a recent
paper~\cite{kls-ofodt-09}, Van Kreveld et al. studied several
types of more complex optimization problems, constrained to
\fodts. They showed that many other criteria can be also optimized
efficiently for $k=1$, making \fodts\ even more appealing for
practical use.

For larger values of $k$, fewer results are known.
The special structure of \fodts\ is not present anymore, which complicates exact optimization algorithms.
Several heuristics and experimental results have been presented for optimization problems related to terrain modeling,
showing that very small values of $k$ ($k=1,\ldots,8$) are enough to achieve important improvements for several terrain criteria~\cite{bg-drthodt-08,bg-sfthodt-08,kkl-grtho-07}.

However, despite the importance given to finding algorithms to
optimize over \hodts, it has never been studied before how many
\hodts\ there can be in the first place. In other words, it is not
known what the minimum and maximum number of different
triangulations are, as functions of $k$ and $n$, not even for the
simpler (but---in practice---most important) case of \fodts.

The problem of determining bounds on the number of \hodts\ is of
both theoretical and practical interest.

From a theoretical point of view, determining how many triangulations a point set has is one of the most intriguing problems in combinatorial geometry, and has received a lot of attention (e.g. \cite{ahhhk-npg-06,sw-rtpps-06,ss-bubnt-03}).
\Hodts\ are a natural and simple generalization of the Delaunay triangulation, hence the impact of such generalization on the number of triangulations is worth studying.

From a more practical point of view, knowing the number of
triangulations for a given $k$ gives an idea of how large the
solution space is when optimizing over this class of well-shaped
triangulations. Ideally, one expects to have many different
triangulations to choose from, in order to find one that is
\emph{good} with respect to other criteria.

Up to now, only trivial bounds were known: every point set has at
least one order-$k$ Delaunay triangulation, for any $k$ (equal to
the Delaunay triangulation), and there are point sets of size $n$
that have $2^{\Theta(n)}$ triangulations, already for $k=1$. In
this paper we present the first non-trivial bounds on the number
of \hodts. Given the practical motivation mentioned above, we are
mostly interested in results that have practical implications for
the use of \hodts. Thus low values of $k$ are our main concern.
Our ultimate goal---achieved partially in this paper---is to
determine to what extent the class of \hodts\ (for small values of
$k$, which has the best triangle-shape properties), also provides
a \emph{large} number of triangulations to choose from.

\paragraph{Results} We study lower and upper bounds on the number of \hodts, as well
as the expected number of order-$k$ Delaunay triangulations for
uniformly distributed points. Let $\maxnok$ denote the maximum
number of order-$k$ Delaunay triangulations that a set with $n$
points can have, and let $\minnok$ denote the minimum number of
order-$k$ Delaunay triangulations that a set with $n$ points can
have. First we show that the lower bound $ \minno{k} \geq 1$ is
tight. In other words, there are arbitrarily large point sets that
have a single \hodt, even for large values of $k$. Next we show
that, for \fodts, $\maxno{1} = 2^{n-3}$. Since these extreme cases
do not describe an average situation when \hodts\ are used, we
then study the number of \hodts\ for a uniformly distributed point
set. Let $R_k$ denote the number of order-$k$ (and not order-$i,$
for any $i < k$) Delaunay triangulations of a uniformly
distributed point set of size $n$. We show that $\bE[R_1]\geq
2^{\rho_1 n(1+o(1))}$, where $\rho_1$ is an analytically defined
constant ($\rho_1 \approx 0.525785$). We also prove that, for
constant values of $k,$ $\bE[R_k]\geq 2^{\rho_k n (1+o(1))}$,
where $\rho_k$ can be calculated numerically (asymptotics are with
respect to $n$). The result has interesting practical
consequences, since it implies that it is reasonable to expect an
exponential number of \hodts\ for any $k \geq 1$.

\paragraph{Related work}
As mentioned earlier, there is no previous work on counting \hodts.
A related concept, the \emph{higher order Delaunay graph}, has been studied by Abellanas et al.~\cite{apgh-sgtph-08}.
The \emph{order-$k$ Delaunay graph} of a set of points $\Pts$ is a graph with vertex set $\Pts$ and an edge between two points $p,q$ when there is a circle through $p$ and $q$ containing at most $k$ other points from $\Pts$.
Abellanas et al. presented upper and lower bounds on the number of edges of this graph. However, since a triangulation that is a subset of the order-$k$ Delaunay graph does not need to be an order-$k$ Delaunay triangulation, it is difficult to derive good bounds for \hodts\ based on them.

There is an ample body of literature on the more general problem of counting \emph{all} triangulations.
Lower and upper bounds on the number of triangulations that $n$ points can have have been improved many times over the years, with the current best ones establishing that there are point sets that have $\Omega(8.48^n)$~\cite{ahhhk-npg-06} triangulations, whereas no point set can have more than $O(43^n)$~\cite{sw-rtpps-06}.

In relation to our expected case analysis of the number of \hodts, it is worth mentioning that many properties of the Delaunay triangulation---and related proximity graphs---of random points have been studied.
The expected behavior of properties of the Delaunay triangulation that have been considered include the average and maximum edge length~\cite{m-hpppp-70,bey-eedt-91a}, the minimum and maximum angles~\cite{bey-eedt-91a}, and its expected weight~\cite{cl-aldt-84}.
Expected properties of other proximity graphs, such as the Gabriel graph and some relatives, are investigated in~\cite{d-essgc-88,c-psepg-92,ms-pggrg-80}.

\paragraph{Outline} This paper is structured as follows.
The next section presents some previous results related to \hodts,
needed for the following sections. In Section~\ref{sec:bounds} we
give lower and upper bounds for the number of \hodts.
Section~\ref{sec:ExpNumTriang} deals with the expected number of
\hodts. Finally, some concluding remarks are made in
Section~\ref{sec:discussion}.

\section{\Hodts}
\label{sec:hodts} We begin by introducing \hodts\ more formally,
and presenting a few properties that will be used throughout the
paper. From now on, we assume that point sets are in general
position.

\begin{definition}\label{def:Order}
(from \cite{ghk-hodt-02}) A triangle $\triangle uvw$ in a point set $\Pts$ is \emph{order-$k$ Delaunay} if its circumcircle $C(u,v,w)$ contains at most $k$ points of $\Pts$. A triangulation of $\Pts$ is \emph{order-$k$ Delaunay} if every triangle of the triangulation is order-$k$.
\end{definition}

Note that if a triangle or triangulation is order-$k$, it is also order-$k'$ for any $k'>k$. A simple corollary of this is that, for any point set and  any $k \geq 0$, the Delaunay triangulation is an order-$k$ Delaunay triangulation.

\begin{definition}
(from \cite{ghk-hodt-02}) An edge $\overline{uv}$ is an order-$k$
Delaunay edge if there exists a circle through $u$ and $v$ that
has at most $k$ points of $\Pts$ inside. An edge $\overline{uv}$
is a \emph{useful} order-$k$ Delaunay edge (or simply
\emph{useful-$k$ edge}) if there is an \orderk\ Delaunay
triangulation that contains $\overline{uv}$.
\end{definition}

\begin{figure}[tb]
\centering
\includegraphics{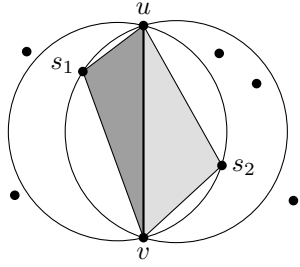}
\caption{The useful order of edge $\overline{uv}$ is determined by
the lowest order of triangles $\triangle uvs_1$ and $\triangle
uvs_2$. In the example, the (lowest) useful order of
$\overline{uv}$ is $\max\{3,1\}=3$.} \label{fig:pre-usefulOrder}
\end{figure}

The useful order of an edge can be checked using the following lemma, illustrated in Figure~\ref{fig:pre-usefulOrder}.

\begin{lemma}
  \label {lem:useful_edge_test}
      (from \cite{ghk-hodt-02}) Let $\overline{uv}$ be an order-$k$ Delaunay edge, let $s_1$ be the point to the left\footnote{
      We sometimes treat edges as directed, to be able to refer to the right or left side of the edge.
      The \emph{left} side of $\overrightarrow{vu}$ denotes the halfplane defined by the line
      supporting $\overline{uv}$, such that
       a polygonal line defined by $v$, $u$ and a point interior to that halfplane, makes a counterclockwise turn.
       In the \emph{right} side, the turn is clockwise.}
       of $\overrightarrow{vu}$, such that the circle $C(u,v,s_1)$ contains no points to the left of $\overrightarrow{vu}$.
       Let $s_2$ be defined similarly but to the right of $\overrightarrow{vu}$. Edge $\overline{uv}$ is
       useful-$k$ if and only if $\triangle{uvs_1}$ and $\triangle{uvs_2}$ are order-$k$ Delaunay triangles.
\end{lemma}

The concept of a \emph{fixed edge} is important in order to study
the structure of \hodts.

\begin{definition}
Let $\Pts$ be a point set and $T$ its Delaunay triangulation. An
edge of $T$ is \emph{$k$-fixed} if it is present in every
order-$k$ Delaunay triangulation of $\Pts$.
\end{definition}

Some simple observations derived from this are that the convex hull edges are always $k$-fixed, for any $k$, and that all the Delaunay edges are $0$-fixed.

\Fodts\ have a special structure.
  If we take all edges that are 1-fixed, then the resulting subdivision has only triangles and
  convex quadrilaterals (and an unbounded face). In the convex quadrilaterals,
  both diagonals are possible to obtain a \fodt\ (see Figure~\ref{fig:FigOrders}, center).
  We say that both diagonals are \emph{flippable}, and similarly we call the quadrilateral \emph{flippable}.
More formally, based on results in~\cite{ghk-hodt-02}, we can make the following observation.

\begin{observation}
\label{obs:Order1Edge}
Let $e$ be a useful order-1 Delaunay edge in an order-1 Delaunay triangulation, such that $e$ is not a Delaunay edge.
Then flipping $e$ results in a Delaunay edge. Moreover, the four edges (different from $e$) that bound the two triangles adjacent to $e$ are 1-fixed edges.
\end{observation}

An implication of this special structure is that instead of
counting triangulations, we can count flippable quadrilaterals or,
equivalently, useful-1 edges that are not Delaunay.

\begin{corollary}
\label{cor:flippable}
Let $\Pts$ be a point set. If $\Pts$ has $q$ flippable quadrilaterals, then $\Pts$ has exactly $2^q$ order-1 Delaunay triangulations.
\end{corollary}

For $k > 1$, the structure is not so simple anymore and it seems
difficult to provide an exact expression for the number of
order-$k$ Delaunay triangulations in terms of the number of
useful-$k$ edges. However, we can derive a lower bound by combining a number of known results, as follows.
First we need some extra
definitions and previous results:

\begin{definition}
(from~\cite{ghk-hodt-02}) The \emph{hull} of an order-$k$ Delaunay
edge $\overline{uv}$ ($k\geq 1$) is the closure of the union of
all Delaunay triangles whose interior intersects $\overline{uv}.$
\end{definition}

\begin{lemma} \label{lem:sizehull}
(from~\cite{ghk-hodt-02}) The hull of an order-$k$ Delaunay edge
($k\geq 1$) is a simple polygon consisting of at most $2k+2$
vertices.
\end{lemma}

\begin{lemma} \label{lem:trianghull}
(from~\cite{ghk-hodt-02}) Let $\overline{uv}$ be a useful-$k$ edge
(and not useful-$i$ for any $i<k$), with $k\geq 1.$ There exists
an order-$k$ (and not order-$i$ for any $i < k$) Delaunay
triangulation of the hull of $\overline{uv}$ that contains
$\overline{uv}.$
\end{lemma}

\begin{lemma} \label{lem:UsefkIntersOrder0}
Let $\overline{uv}$ be an order-$0$ edge. The number of useful-$k$
edges ($k\geq 1$) that intersect $\overline{uv}$ is at most
$(2k+1)^2.$
\end{lemma}

\begin{proof}
It follows directly from the proof of Lemma~8
in~\cite{ghk-hodt-02}.
\end{proof}

We have now the necessary tools to prove the following lower bound on the number of order-$k$ triangulations, expressed as a function of the number of useful-$k$ edges.

\begin{lemma}\label{lem:k_flippable}
Let $\Pts$ be a point set and let $e_k,$ for $k>1,$ be the number
of useful-$k$ edges (which are not useful-$i$ for any $i<k$) of
$\Pts.$ Then $\Pts$ has at least $2^{e_k/C_k}-1$ order-$k$ (and
not order-$i,$ for any $i < k$) Delaunay triangulations, where
$C_k=(4k+1)(2k+1)^2+1.$
\end{lemma}

\begin{proof}
Let $E_k$ denote the set of useful-$k$ edges (which are not
useful-$i$ for any $i<k$) of $\Pts$ and let $e_k$ denote the
cardinal number of $E_k.$ We select a subset $E'_k$ of the edges
of $E_k$ in the following way: We pick an edge $e$ of $E_k,$ we
remove all the edges in $E_k$ whose hull intersects the hull of
$e$ in at least one Delaunay triangle, and we repeat until $E_k$
does not contain any edge.

Let $e'$ be an edge in $E'_k$ and $e$ be an edge in $E_k$ whose
hull intersects the hull of $e'$ in at least one Delaunay triangle
$T.$ Then $e$ intersects at least one edge of $T.$ By
Lemma~\ref{lem:sizehull}, all Delaunay triangles included in the
hull of $e'$ contain at most $4k+1$ edges. By
Lemma~\ref{lem:UsefkIntersOrder0}, each of them intersects at most
$(2k+1)^2$ useful-$k$ edges. Hence if $e'$ is selected, at most
$(4k+1)(2k+1)^2$ edges in $E_k$ are removed. Therefore $E'_k$
contains at least $e_k/((4k+1)(2k+1)^2+1)$ edges.

Each non-empty subset of $E'_k$ gives rise to a different
order-$k$ (and not order-$i,$ for any $i < k$) Delaunay
triangulation proceeding as follows: If an edge $e$ is in the
subset, we triangulate the hull of $e$ as in
Lemma~\ref{lem:trianghull}, that is, we use an order-$k$ (and not
order-$i,$ for any $i < k$) Delaunay triangulation containing $e.$
If an edge $e$ is not in the subset $E'_k$, we triangulate the hull of
$e$ using the Delaunay triangles crossed by $e.$
Finally, we complete the
triangulation by adding Delaunay triangles in the regions that
have not been triangulated (that is, computing a constrained Delaunay
triangulation). This construction is consistent because the hulls
of the edges in $E'_k$ can only intersect in points and boundary
edges, and because the boundary edges of the hulls belong to the
Delaunay triangulation.
\end{proof}

\section{Lower and upper bounds}
\label{sec:bounds}

In this section we derive upper and lower bounds on the number of \fodts.
As mentioned in the introduction, due to the practical motivation of this work, we are mostly interested in lower bounds.
However, for completeness and because the theoretical question is also interesting, this section also includes a result on upper bounds.

The main question that we address in this section is: what is the minimum number of \hodts\ that $n$ points can have?
Are there arbitrarily large point sets that have only $O(1)$ \hodts?
To our surprise, the answer to the second question is affirmative.

The lemma below presents a construction that has only one \hodt, regardless of the value of $k$, for any $k \leq \left\lfloor n/3 \right\rfloor-1$.
Note that this implies that for any value of $k$ of practical interest, there are point sets that have no other order-$k$ Delaunay triangulation than the Delaunay triangulation.

\begin{figure}[tb]
\centering
\includegraphics{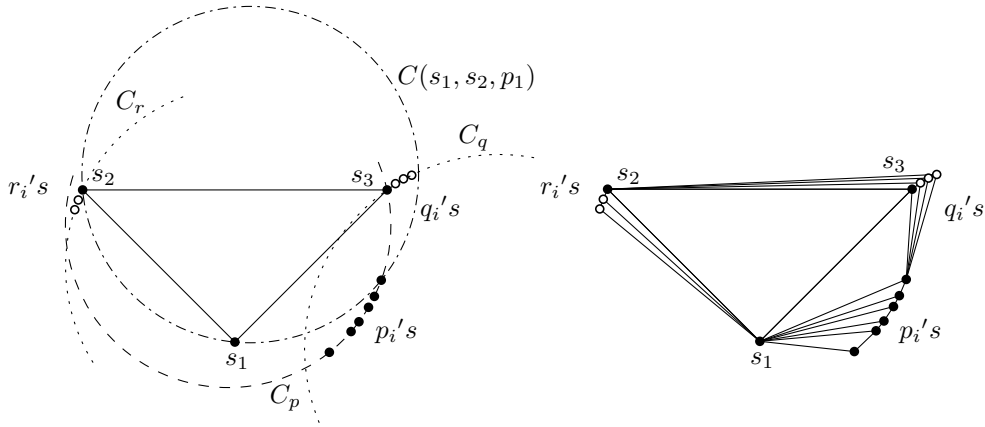}
\caption{Construction of a point set (left) whose only order-$k$ Delaunay triangulation is the Delaunay triangulation (right).}
\label{fig:ExampleOnly1Triangulation}
\end{figure}

\begin{lemma}
\label{lem:Only1}
Given any $n \geq 6$ and any $k$ such that $k \leq \left\lfloor n/3\right\rfloor-1$, there are point sets with $n$ points in general position that have only one order-$k$ Delaunay triangulation.
\end{lemma}

\begin{proof}
We give a construction with $n$ points that can be shown to have only one order-$k$ Delaunay triangulation, for any value of  $k \leq \left\lfloor n/3\right\rfloor -1$. The construction is illustrated in Figure~\ref{fig:ExampleOnly1Triangulation}.

To simplify the explanation, in the following we assume that $n$ is a multiple of 3.
Since any order-$i$ Delaunay triangulation is also order-$k$ for all $k \geq i$, for the proof it is enough to use $k = n/3 -1$.

We start with a triangle $\triangle s_1 s_2 s_3$. Then we add
three groups of points, which we will denote with letters $p$,
$q$, and $r$. The points in the first group are
denoted $p_1 \dots p_m$, where $m=n/3$. These points are initially
placed on a circle $C_p$ that goes through $s_3$, as shown in the
figure; they are sorted from top to bottom. The second group
comprises $k$ points $q_1 \dots q_k$, placed very close to each
other on a circle $C_q$ through $s_3$, as shown in the figure. In
addition, we must also make sure that the points $q_1 \dots q_k$
are close enough to $s_3$ in order to be contained inside
$C(s_1,s_2,p_1)$. Finally, the points in the third group, $r_1
\dots r_{k-1}$, are placed very close to each other on a third
circle $C_r$, which goes through $s_2$. The important properties
of these circles are:
(i) $C_p$ contains $s_2$ and all the points $r_i$,
(ii) $C_r$ and $C_q$ contain all the points of the type $p_i$,
and
(iii) $C(s_1,s_2,p_1)$ contains $s_3$ and all points of type $q_i$.

Clearly, the point set as constructed is degenerate, but this can be easily solved by applying a slight perturbation to each point, without affecting the properties just mentioned.
Moreover, the perturbation can be made such that the Delaunay triangulation of the point set looks like the one in the right of Figure~\ref{fig:ExampleOnly1Triangulation}.

We now argue that all the edges in the Delaunay triangulation are
$k$-fixed, by considering the different types of edges that,
potentially, could cross a Delaunay edge to make it non-fixed.
Suppose an edge of the shape $\overline{s_1 p_i}$ is not
$k$-fixed. Then there must be some triangulation in which the edge
is crossed by some other useful order-$k$ edge. Such edge can be
of three types: (i) it connects two points $p_j$, $p_k$, (ii) it
connects two points $p_j$, $q_k$ (or $s_3$), or (iii) it connects
two points $p_j$, $r_k$ (or $s_2$). An edge of the type
$\overline{p_j p_k}$ that crosses $\overline{s_1 p_i}$ must be an
edge of the shape $\overline{p_ip_{i+2}}$ or force a such an edge
to appear in the triangulation. However, the circumcircle of the
triangle defined by any three consecutive points $p_i$, $p_{i+1}$,
$p_{i+2}$ contains at least $k+1$ points because it is a slightly
perturbed version of $C_p$. Thus no such edge can be part of an
order-$k$ triangulation. A similar situation occurs with any edge
of the shape $\overline{p_j q_k}$, since it forces a triangle of
the form $\triangle p_ip_lq_m$ (or $\triangle p_ip_ls_3$).
Finally, edges of type $\overline{p_j r_k}$ force a triangle of
the form $\triangle s_1p_ir_l$ (or $\triangle s_1p_is_2$), whose
circumcircle includes at least as many points as contained in
$C(s_1,s_2,p_1)$, hence cannot be part of an order-$k$
triangulation either. Therefore all the edges of the shape
$\overline{s_1 p_i}$ are $k$-fixed. Similar arguments can be used
to show that the edges in the other groups are also $k$-fixed,
hence no other order-$k$ triangulation can exist.
\end{proof}

Having determined that some point sets can have as little as one \fodt, it is reasonable to ask what is the \emph{maximum} number of \fodts\ that a point set can have.
The following lemma gives a precise---and tight---bound on the maximum number of \fodts.

\begin{figure}[tb]
\centering
\includegraphics{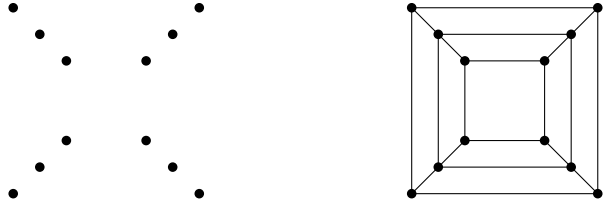}
\caption{Construction achieving the maximum number of \fodts. Left
and right: point set and flippable quadrilaterals, for points not
in general position.} \label{fig:maxNumberOrder1bis}
\end{figure}

\begin{lemma} \label{lem:maxfodt}
Every point set $\Pts$ with $n$ points in general position has at most
$2^{n-3}$ \fodts, and this bound is tight.
\end{lemma}
\begin{proof}
To see that no point set can have more than $n-3$ flippable
quadrilaterals, observe that the subdivision of the convex hull of
$\Pts$ induced by the fixed edges is a plane graph. It follows
from Euler's equation that any triangulation has at most $2n-5$
triangles. Since each quadrilateral is formed by two triangles,
there can be at most $n-3$ quadrilaterals.

Now we give a construction with $n$ (for $n$ any multiple of 4)
points that has $n-3$ flippable quadrilaterals, thus a total of
$2^{n-3}$ \fodts. The construction is illustrated in
Figure~\ref{fig:maxNumberOrder1bis}, and consists of a series of
points placed on the vertices of concentric squares with the same
orientation. Clearly, the edges in
Figure~\ref{fig:maxNumberOrder1bis} are Delaunay edges and the
four vertices of each quadrilateral are cocircular. If we apply a
small perturbation to the point set so that it reaches a general
position, one of the diagonals of each quadrilateral becomes a
Delaunay edge, while the other one becomes a useful order-1
(non-Delaunay) edge. Therefore, all quadrilaterals are flippable.
\end{proof}

\section{Expected number of triangulations}
\label{sec:ExpNumTriang} Let $\Pts$ be a set of $n$ points
uniformly distributed in the unit square. In this section we give
lower bounds on the expected number of \hodts\ of $\Pts.$

Note  that the events that four points in $\Pts$
are cocircular and that three points in $\Pts$ form a right angle
happen with probability zero, and hence we can safely ignore these cases. Throughout this section we will use the notation $x \sim y$ if $x=y(1+o(1))$.

We start with \fodts. We aim to compute the probability that two
randomly chosen points $u, v$ in $\Pts$ form a useful-1,
non-Delaunay, edge. Assume that the edge is directed
$\overrightarrow{vu}$. Let $w$ be the point to the left of
$\overrightarrow{vu}$, such that the circle $C(u,v,w)$ contains no
points to the left of $\overrightarrow{vu}$, and let $t$ be the
point to the right of $\overrightarrow{vu}$, such that the circle
$C(u,v,t)$ contains no points to the right of
$\overrightarrow{vu}$. Let $\mathcal{E}$ be the event defined as
follows: the edge $\overline{uv}$ is useful-1 (but not Delaunay),
$t$ is to the right of $\overrightarrow{vu}$ and the circle
$C(u,v,t)$ contains no points of $\Pts$ to the right of
$\overrightarrow{vu}$, $w$ is to the left of $\overrightarrow{vu}$
and the circle $C(u,v,w)$ contains no points of $\Pts$ to the left
of $\overrightarrow{vu}$. It is well-known that $\overline{uv}$
belongs to the Delaunay triangulation of $\Pts$ if and only if
$\angle uwv+\angle utv < \pi$. Thus the event $\mathcal{E}$ can be
decomposed into the disjoint union
$\mathcal{E}=\mathcal{E}_1\cup\mathcal{E}_2\cup\mathcal{E}_3,$
where $\mathcal{E}_1$ denotes the event $\mathcal{E}$ with the
additional conditions that $\angle uwv > \pi/2,$ $\angle utv >
\pi/2,$ $\mathcal{E}_2$ denotes the event $\mathcal{E}$ with the
 conditions that $\angle uwv< \pi/2,$ $\angle utv >
\pi/2,$ and $\mathcal{E}_3$ denotes the event $\mathcal{E}$ with
the  conditions that $\angle uwv> \pi/2,$ $\angle utv <
\pi/2$ (in all cases we must have $\angle uwv+\angle utv
>\pi$).  Consequently,
$\bP[\mathcal{E}]=\bP[\mathcal{E}_1]+\bP[\mathcal{E}_2]+\bP[\mathcal{E}_3].$

\begin{lemma}\label{lem_e1_e2}
$\bP[\mathcal{E}_1]\sim c_1/n^3$ and $\bP[\mathcal{E}_2] \sim c_2/n^3$, where $c_1 \stackrel{\cdot}{=} 0.23807$ and $c_2 \stackrel{\cdot}{=} 0.40675$.
\end{lemma}
\begin{proof}
Let us first compute $\bP[\mathcal{E}_1].$

Let $A_w$ (respectively, $A_t$)  be the interior of the set
consisting of all points in $C(u,v,w)\cap C(u,v,t)$ that are to
the left (resp. right) of $\overrightarrow{vu}.$ Let $B_w$
(respectively, $B_t$) denote the interior of the set containing
all points in $C(u,v,w)$ (resp. $C(u,v,t)$) that are to the right
(resp. left) of $\overrightarrow{vu}$ and do not lie in $A_t$
(resp. $A_w$) (see Figure~\ref{fig:Events}, left).
 Since $w$ is the point such that the
circle $C(u,v,w)$ contains no points to the left of
$\overrightarrow{vu},$ the region $A_w$ is empty of points in
$\Pts.$ In order for the edge $\overline{uv}$ to be useful-1, the region $B_t$ also has to be empty of points in $\Pts.$
Analogously, under the hypothesis of $\mathcal{E}_1$, the regions
$A_t$ and $B_w$ are empty of points in $\Pts.$ It is not
difficult to see that the reverse implications also hold.
Therefore, the event $\mathcal{E}_1$ is equivalent to the event
that $A_w,$ $B_t,$ $A_t,$ and $B_w$ do not contain any point in
$\Pts.$

\begin{figure}[tb]
\centering \scalebox{0.82}{\input{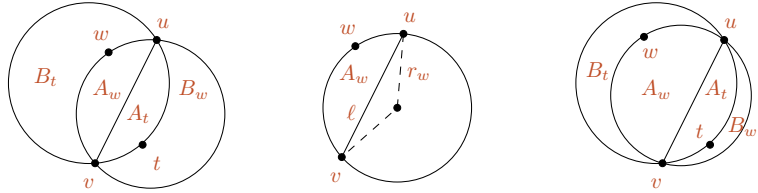}}
\caption{Left: the event $\mathcal{E}_1$ and the regions $A_w$,
$B_w$, $A_t$ and $B_t$. Middle: in the event $\mathcal{E}_1$, the
region $A_w$ is a circular sector minus a triangle. Right: the
event $\mathcal{E}_2$ and the regions $A_w$, $B_w$, $A_t$ and
$B_t$.} \label{fig:Events}
\end{figure}

Now let us denote by $r_w$ the radius of the circle $C(u,v,w)$ and
by $\ell$ the length of the edge $\overline{uv}$ (see
Figure~\ref{fig:Events}, center). A straightforward calculation
leads to the following expression for the area of $A_w$:
\begin{eqnarray*}
\textrm{area}\,(A_w)=r_w^2\arcsin\left(\frac{\ell}{2r_w}\right)-\frac{\ell}{2}\sqrt{r_w^2-\frac{\ell^2}{4}}.
\end{eqnarray*}

In order to compute $\bP[\mathcal{E}]$, we will be interested in
having certain areas being empty of $n$ points, which happens with
probability $(1-\textrm{area}\,(A))^{n}$ (for $A$ the area in
question). Since the contribution of areas $A$ of size $\Theta(1)$
is $O(\lambda^n)$ for some $0 < \lambda < 1$ (which is far less
than the asymptotic value of the integrals, as we shall see
below), for any constant $j$ we can safely assume in the
integrals below the asymptotic equivalence
$(1-\textrm{area}\,(A))^{n-j} \sim e^{-\textrm{area}\,(A)n}$,
without affecting the first order terms of the asymptotic behavior
of the integral~\footnote{In fact, this formula arises in a
homogeneous Poisson point process of intensity $n$ in the unit
square, and it is not surprising that both distributions give the
same asymptotic results (see the ideas of depoissonization given
in~\cite{Penrose03}).}.

Observe that $\ell$ may take values from $0$ to $\sqrt{2}$ and
that the probability density of the event $|\overline{uv}|=\ell$
is $2\pi \ell\,d\ell$. Notice also that
$r_w \in (\ell/2,+\infty)$ and that the event of having a radius $r_w$
has probability density
$\left(-2r_w\arcsin\left(\frac{\ell}{2r_w}\right)+\frac{\ell r_w}{\sqrt{r_w^2-\frac{\ell^2}{4}}}\right)dr_w,$
since it corresponds to the negative derivative $-f'(r_w) dr_w$ of the function
$f(r)=r^2\arcsin\left(\frac{\ell}{2r}\right)-\frac{\ell}{2}\sqrt{r^2-\frac{\ell^2}{4}}.$

Denoting by $r_t$ the radius of the circle $C(u,v,t)$, we obtain analogous expressions for $r_t$.

Now we have all the necessary ingredients to develop an expression
for $\bP[\mathcal{E}_1]:$

\begin{equation}\label{eq:case1_formula}
\begin{split}
\bP[\mathcal{E}_1] & \sim \int_0^{\sqrt{2}} 2\pi \ell
\int_{\ell/2}^{\infty} \int_{\ell/2}^{\infty}
\left(-2r_w\arcsin\left(\frac{\ell}{2r_w}\right)+\frac{\ell
r_w}{\sqrt{r_w^2-\frac{\ell^2}{4}}}\right) \\ & \quad \quad
\left(-2r_t\arcsin\left(\frac{\ell}{2r_t}\right)+\frac{\ell r_t}{\sqrt{r_t^2-\frac{\ell^2}{4}}}\right) \\
& \quad \quad e^{-n(\pi r_w^2+\pi r_t^2
-r_w^2\arcsin(\frac{\ell}{2r_w})+\frac{\ell}{2}\sqrt{r_w^2-\frac{\ell^2}{4}}
-r_t^2\arcsin(\frac{\ell}{2r_t})+\frac{\ell}{2}\sqrt{r_t^2-\frac{\ell^2}{4}})}dr_tdr_wd\ell
\end{split}
\end{equation}
Since classical methods for asymptotic integration fail for the
integral given by~\eqref{eq:case1_formula} (the derivative of the
exponent is infinity at the point where the exponent maximizes),
we apply the following change of variables:  $\ell/2= a/\sqrt n$,
$r_t=b/\sqrt n$, $r_w = c/\sqrt n$. The
integral~\eqref{eq:case1_formula} then becomes (replacing the
integration limit $\sqrt{2n}$ by $\infty$, which can be done since
the dominant contribution comes from small values of $a$)

\begin{eqnarray*}\label{eq:case1_2}
 \bP[\mathcal{E}_1]&\sim&\frac{1}{n^3}
\int_0^{\infty} \int_{a}^{\infty} \int_{a}^{\infty} 8 \pi a
\left(-2c \arcsin\left(\frac{a}{c}\right)+\frac{2 ac}{\sqrt{c^2-a^2}} \right) \notag\\
&& \left(-2b \arcsin\left(\frac{a}{b}\right)+\frac{2 ab}{\sqrt{b^2-a^2}} \right)  \notag\\
&& e^{-\pi b^2 + b^2\arcsin\left(\frac{a}{b}\right) -a
\sqrt{b^2-a^2}-\pi c^2+c^2 \arcsin\left(\frac{a}{c}\right)-a
\sqrt{c^2-a^2} } db\,dc\,da.
\end{eqnarray*}

Given that it does not seem possible to evaluate this integral
symbolically, we resort to applying numerical methods. For reasons
of numerical stability (especially in the case of
$\bP[\mathcal{E}_2]$ below) we apply another change of variables
$b=\frac{a}{\sin(\sigma/2)}$, $c=\frac{a}{\sin(\theta/2)}$:

\begin{eqnarray*}\label{eq:case1_3}
 \bP[\mathcal{E}_1]&\sim&\frac{1}{n^3}
\int_0^{\infty} \int_{0}^{\pi} \int_{0}^{\pi} 8 \pi a \left(
\frac{a^2 \theta \cot(\theta/2)-2a^2}{1-\cos(\theta)} \right)
\left( \frac{a^2 \sigma \cot(\sigma/2)-2a^2}{1-\cos(\sigma)}
\right)
 \notag\\
&& e^{a^2(\frac{\theta-2\pi}{1-\cos(\theta)} - \cot(\theta/2)+
\frac{\sigma-2\pi}{1-\cos(\sigma)} - \cot(\sigma/2))} d\sigma
d\theta da.
\end{eqnarray*}

Solving this integral numerically (using Mathematica), we obtain
that $\bP[\mathcal{E}_1]\sim c_1/n^3$, where $c_1
\stackrel{\cdot}{=} 0.23807$.

Let us now consider $\mathcal{E}_2.$ Let $A_w$, $B_w$, $A_t$ and
$B_t$ be defined as in the event $\mathcal{E}_1$ (see
Figure~\ref{fig:Events}, right). By the same arguments, the event
$\mathcal{E}_2$ is equivalent to the event that the regions $A_w,$
$B_t,$ $A_t,$ and $B_w$ are empty of points in $\Pts.$ Analogous
observations as in the previous case yield

\begin{eqnarray}\label{eq:case2_formula}
 \bP[\mathcal{E}_2]&\sim&
\int_0^{\sqrt{2}} 2\pi \ell
\int_{\ell/2}^{\infty}\int_{r_w}^{\infty}
\left(2r_w \pi -2r_w \arcsin\left(\frac{\ell}{2r_w}\right)+\frac{\ell r_w}{\sqrt{r_w^2 - \frac{\ell^2}{4}}} \right) \notag\\
&& \left(-2r_t \arcsin\left(\frac{\ell}{2r_t}\right)+\frac{\ell r_t}{\sqrt{r_t^2 - \frac{\ell^2}{4}}} \right) \notag\\
&& e^{-n(\pi r_t^2 - r_t^2
\arcsin\left(\frac{\ell}{2r_t}\right)+\frac{\ell}{2}\sqrt{r_t^2-\frac{\ell^2}{4}}+r_w^2
\arcsin\left(\frac{\ell}{2r_w}\right)-\frac{\ell}{2}\sqrt{r_w^2-\frac{\ell^2}{4}}
)} dr_t dr_w d\ell.
\end{eqnarray}

As before, we apply the substitution $\ell/2= a/\sqrt n$,
$r_t=b/\sqrt n$, $r_w = c/\sqrt n$ to the
integral~\eqref{eq:case2_formula} and obtain

\begin{eqnarray*}\label{eq:case2_2}
  \bP[\mathcal{E}_2]&\sim&\frac{1}{n^3}
\int_0^{\infty} \int_{a}^{\infty} \int_{c}^{\infty} 8 \pi a
 \left(2c\pi-2c \arcsin\left(\frac{a}{c}\right)+\frac{2 ac}{\sqrt{c^2-a^2}} \right)  \notag\\
&& \left(-2b \arcsin\left(\frac{a}{b}\right)+\frac{2 ab}{\sqrt{b^2-a^2}} \right) \notag\\
&& e^{-\pi b^2 + b^2\arcsin\left(\frac{a}{b}\right) -a
\sqrt{b^2-a^2}-c^2 \arcsin\left(\frac{a}{c}\right)+a
\sqrt{c^2-a^2} } db\,dc\,da.
\end{eqnarray*}

For reasons of numerical stability, we again apply the change of
variables $b=\frac{a}{\sin(\sigma/2)}$ and
$c=\frac{a}{\sin(\theta/2)}$ and obtain
\begin{eqnarray*}\label{eq:case2_3}
 \bP[\mathcal{E}_2]&\sim&\frac{1}{n^3}
\int_0^{\infty} \int_{0}^{\pi} \int_{0}^{\theta} 8 \pi a \left(
\frac{a^2 (\theta -2\pi)\cot(\theta/2)-2a^2}{1-\cos(\theta)}
\right) \left( \frac{a^2 \sigma
\cot(\sigma/2)-2a^2}{1-\cos(\sigma)} \right)
 \notag\\
&& e^{a^2(\cot(\theta/2)-\frac{\theta}{1-\cos(\theta)}+
\frac{\sigma-2\pi}{1-\cos(\sigma)} - \cot(\sigma/2))} d\sigma
d\theta da.
\end{eqnarray*}

Solving this integral numerically (using Mathematica), we obtain
that  $\bP[\mathcal{E}_2]\sim c_2/n^3$, where $c_2
\stackrel{\cdot}{=} 0.40675$.
\end{proof}

Denote by $U_1$ the random variable counting the number of
useful-1 (and not Delaunay) edges. We have the following
corollary:
\begin{corollary}\label{cor:useful_1}
$\bE[U_1] \sim\frac{c_1+2c_2}{2n} =: d_1 n,$ where $d_1
\stackrel{\cdot}{=} 0.525785$.
\end{corollary}
\begin{proof}
Since $\mathcal{E}_2$ and $\mathcal{E}_3$ are symmetric, we
obviously have that $ \bP[\mathcal{E}_3]=\bP[\mathcal{E}_2].$
Hence, $\bP[\mathcal{E}]\sim \frac{c_1+2c_2}{n^3}.$ Since for a
fixed edge $\overline{uv}$ there are $(n-2)(n-3) \sim n^2$ ways to
choose the points $w$ and $t$ to the left and to the right of
$\overrightarrow{vu}$, and these events are all disjoint, the edge
$\overline{uv}$ is useful-1 (and not Delaunay) with probability
$\frac{c_1+2c_2}{n}$. Hence, $\bE[U_1] \sim \binom{n}{2}
\frac{c_1+2c_2}{n} \sim \frac{c_1+2c_2}{2n} =: d_1 n,$ where $d_1
\stackrel{\cdot}{=} 0.525785$.
\end{proof}

Recall that $R_k$ denotes the number of order-$k$ (and not
order-$i,$ for any $i < k$) Delaunay triangulations of a uniformly
distributed point set. We can now state the following theorem:

\begin{theorem}\label{thm:triangulations1}
Given $n$ points distributed uniformly at random in the unit
square, $\bE[R_1] \geq 2^{\rho_1 n(1+o(1))}$, where $\rho_1
\stackrel{\cdot}{=} 0.525785.$
\end{theorem}
\begin{proof}
By Corollary~\ref{cor:flippable}, $\bE[R_1]=\bE[2^{U_1}]$. Now, by
Jensen's inequality, $\bE[2^{U_1}] \geq 2^{\bE[U_1]}$, and the
result follows by Corollary~\ref{cor:useful_1}.
\end{proof}

Combining the ideas for the case $k=1$ with the result from
Lemma~\ref{lem:k_flippable}, we obtain the following
generalization for constant values of $k$:

\begin{theorem}
Given $n$ points distributed uniformly at random in the unit
square, for any constant value of $k,$ $\bE[R_k] \geq 2^{\rho_k
n(1+o(1))},$ where $\rho_k$ is a constant that can be calculated
numerically.
\end{theorem}
\begin{proof}
Denote by $U_k$ the number of useful-$k$ edges (which are not
useful-$i$ for any $i<k$) for any constant $k
> 1$. We want to know the value of $\bE[U_k]$.

In order for an edge $\overline{uv}$ to be useful-$k$ (and not
useful-$i$ for $i=0,\ldots,k-1$), using the notation of
Figure~\ref{fig:Events}, first observe that the regions $A_w$ and
$A_t$ have to be empty of points. Moreover, either the region $B_t
\setminus A_w$ has to contain exactly $k-1$ points ($w$ is
excluded), whereas the region $B_w \setminus A_t$ can contain any
number of points $i=0,\ldots,k-1$ ($t$ is excluded), or vice
versa. For any constant $i$, the probability of having exactly $i$
points in an area $A$ of size $o(1)$ (as before, for constant $i$
only such areas count for the asymptotic behaviour of the
integrals) is $\sim e^{-n\textrm{A}} (n\textrm{A})^i/i!$. Thus,
defining the events $\mathcal{E}_1,$ $\mathcal{E}_2,$ and
$\mathcal{E}_3$ analogously as in Section~\ref{sec:ExpNumTriang},

\begin{eqnarray*}\label{eq:case1_k}
 \bP[\mathcal{E}_1]&\sim&
\sum_{i=0}^{k-1} \int_0^{\sqrt{2}} 2\pi \ell
\int_{\ell/2}^{\infty}\int_{\ell/2}^{\infty}
\left(-2r_w \arcsin\left(\frac{\ell}{2r_w}\right)+\frac{\ell r_w}{\sqrt{r_w^2 - \frac{\ell^2}{4}}} \right) \notag\\
&& \left(-2r_t \arcsin\left(\frac{\ell}{2r_t}\right)+\frac{\ell r_t}{\sqrt{r_t^2 - \frac{\ell^2}{4}}} \right) \notag\\
&& e^{-n(\pi r_w^2+\pi r_t^2 - r_w^2
\arcsin\left(\frac{\ell}{2r_w}\right)
+\frac{\ell}{2}\sqrt{r_w^2-\frac{\ell^2}{4}}-r_t^2
\arcsin\left(\frac{\ell}{2r_t}\right)+\frac{\ell}{2}\sqrt{r_t^2-\frac{\ell^2}{4}}
)} \notag\\
&& (\textrm{area}\,(B_t \setminus A_w)\,n)^{k-1}
(\textrm{area}\,(B_w \setminus A_t)\,n)^{i} \frac{1}{(k-1)!i!}
dr_t dr_w d\ell \notag\\
+ && \sum_{i=0}^{k-2} \int_0^{\sqrt{2}} 2\pi \ell
\int_{\ell/2}^{\infty}\int_{\ell/2}^{\infty}
\left(-2r_w \arcsin\left(\frac{\ell}{2r_w}\right)+\frac{\ell r_w}{\sqrt{r_w^2 - \frac{\ell^2}{4}}} \right) \notag\\
&& \left(-2r_t \arcsin\left(\frac{\ell}{2r_t}\right)+\frac{\ell r_t}{\sqrt{r_t^2 - \frac{\ell^2}{4}}} \right) \notag\\
&& e^{-n(\pi r_w^2+\pi r_t^2 - r_w^2
\arcsin\left(\frac{\ell}{2r_w}\right)
+\frac{\ell}{2}\sqrt{r_w^2-\frac{\ell^2}{4}}-r_t^2
\arcsin\left(\frac{\ell}{2r_t}\right)+\frac{\ell}{2}\sqrt{r_t^2-\frac{\ell^2}{4}}
)} \notag\\
&& (\textrm{area}\,(B_t \setminus A_w)\,n)^{i}
(\textrm{area}\,(B_w \setminus A_t)\,n)^{k-1}  \frac{1}{(k-1)!i!}
dr_t dr_w d\ell.
\end{eqnarray*}

Now, since
\begin{eqnarray*}
\textrm{area}\,(B_t \setminus A_w)=r_t^2 \pi-r_t^2 \arcsin\left(\frac{\ell}{2r_t}\right)
+\frac{\ell}{2}\sqrt{r_t^2-\frac{\ell^2}{4}}-r_w^2
\arcsin\left(\frac{\ell}{2r_w}\right)+\frac{\ell}{2}\sqrt{r_w^2-\frac{\ell^2}{4}}
\end{eqnarray*}
and
\begin{eqnarray*}
\textrm{area}\,(B_w \setminus A_t)=r_w^2 \pi-r_w^2 \arcsin\left(\frac{\ell}{2r_w}\right)
+\frac{\ell}{2}\sqrt{r_w^2-\frac{\ell^2}{4}}-r_t^2
\arcsin\left(\frac{\ell}{2r_t}\right)+\frac{\ell}{2}\sqrt{r_t^2-\frac{\ell^2}{4}},
\end{eqnarray*}
after applying the substitutions $\ell/2= a/\sqrt n$, $r_t=b/\sqrt
n$, $r_w = c/\sqrt n$, in these new factors $n$ disappears and the
integral again yields $\Theta(1/n^3)$. The same argument also
holds for $\bP[\mathcal{E}_2],$ and by the same reasoning as in
the case of useful-1 edges, we obtain that the expected number of
useful-$k$ edges (that are not useful-$i$ for any $i < k$) is $d_k
n$ for any constant $k$. We point out that using our formula the
constant $d_k$ can be calculated numerically.

By Lemma~\ref{lem:k_flippable}, $R_k\geq 2^{U_k/C_k}-1,$ where
$C_k=(4k+1)(2k+1)^2+1.$ Therefore $\bE[R_k]\geq
\bE[2^{U_k/C_k}]-1$ and as before, by Jensen's inequality,
$\bE[2^{U_k/C_k}]\geq 2^{\bE[U_k]/C_k}.$
\end{proof}

\section{Discussion and further work}
\label{sec:discussion} We have given the first non-trivial bounds
on the number of \hodts. We showed that there are sets of $n$
points that have only one \hodt\ for values of $k \leq
\left\lfloor n/3 \right\rfloor-1$, and that no point set can have
more than $2^{n-3}$ \fodts. Moreover, we showed that for any
constant value of $k$ (in particular, already for $k=1$) the
expected number of order-$k$ triangulations of $n$ points
distributed uniformly at random is exponential. This supports the
use of \hodts\ for small values of $k$, which had already been
shown to be useful in several applications related to terrain
modeling~\cite{kkl-grtho-07}. From a more theoretical perspective,
it would be interesting to obtain tighter bounds on the expected number of
order-$k$ Delaunay triangulations for uniformly distributed points.
\paragraph{Acknowledgments} We are grateful to Nick Wormald for his help on dealing with the integrals of Section~\ref{sec:ExpNumTriang}. We would also like to thank Christiane Schmidt and Kevin Verbeek for helpful discussions.
M. S. was partially supported by projects MTM2009-07242 and Gen. Cat. DGR
2009SGR1040.
R.I.S. was supported by the Netherlands Organisation for
  Scientific Research (NWO).

\bibliographystyle{abbrv}

\end{document}